# BİLGİSAYAR MÜHENDİSLİĞİ EĞİTİMİNDE TEKNOLOJİ EĞİLİMLERİNİN TAKİP EDİLMESİ

**Ahmet Murat Türk [1], Alper Bilge [2]**

## Abstract

Günümüzde bilgisayarlar yaşamın ayrılmaz bir parçası haline gelmiştir. Ancak pek çok kimsenin bilgisayarla olan etkileşimi son kullanıcı düzeyindedir. Bilgisayar sistemlerinin yapısı, tasarımı, geliştirilmesi ve sistemlerin kullanılması, ihtiyaç halinde gerekli yazılımların geliştirilmesi, donanım ve yazılım sorunlarının çözülmesinde bilgisayar mühendislerine ihtiyaç duyulmaktadır. Yetkin bilgisayar mühendislerinin yetiştirilmesi, gelecek teknolojilerinde söz sahibi olmak için önem arz etmektedir. Bilgisayar teknolojilerinin eğilimleri sürekli değişmektedir. 1990'lı yıllarda bilgisayar sistemlerindeki çalışmalar taşınabilirlik, doğruluk, çalışma süresi, verimlilik gibi metrikleri iyileştirmeye yoğunlaşmışken, 2000'li yıllarla güvenlik, sürdürülebilirlik, ölçeklenebilirlik, kişisel gizlilik gibi konular önemli ölçütler haline gelmiştir. Son dönemlerde büyük veri analizi, bulut teknolojileri, giyilebilir ve çevrimiçi hizmetlerin mobil uyumları gibi konular popülerdir. Buna karşılık bilgisayar mühendisliği eğitimi de kendini güncellemeli ve çağa ayak uydurmalıdır. Bu çalışmada, yakın zamanda yaygınlaşması beklenen ve bilgisayar mühendisliği öğrencilerinin yetkinlik kazanması gerektiği; büyük veri, giyilebilir cihazlar, nesnelerin interneti, bulut teknolojileri, kimlik yönetimi ve siber güvenlik gibi alanlarda bilgisayar mühendisliği müfredatlarına eklenmesi önerilen konulara değinilmiştir. İlgili konuların tanımları yapılarak kullanım alanları açıklanmış, gelişimi ve gelecekte oynayacağı rollere değinilmiş, konuların kapsamları tarif edilerek müfredatta yer alması önerilen bölümleri ve elde edilecek kazanımlar açıklanmıştır. Ayrıca bu kazanımların MÜDEK akreditasyonu sürecinden geçmiş bilgisayar mühendisliği bölümlerinin öğrenim çıktıları ile ilişkileri belirtilmiştir.

# FOLLOWING TECHNOLOGY TRENDS IN COMPUTER ENGINEERING EDUCATION

## Abstract

Today computers has become an integral part of life. However, most people's interaction with computers in on end-user-level. Computer Engineers are needed while designing and developing structure of computer systems, software and hardware systems and also they are need when implementing and solving problems while using these systems. Training of qualified computer engineers are vital to have a say in future technology. Computer technology is constantly changing trends. Studies on computer systems are focused on improving portability, accuracy, run time in 1990s, yet metrics such as productivity, security sustainability, scalability, and privacy issues becomes important criteria in 2000s. Recently, big data analysis, cloud technologies, wearable technologies, mobile and online services are become popular. For that reason, computer engineering education should be update itself regularly and keep up with latest improvements. In this study, it is touched on some topics which is suggested to extend computer engineering curriculum such as big data analyses, wearable technologies internet of things, cloud technologies, identity management and cyber security which are expected to widening in the area and also demanded that computer engineering student should be qualified on. Related topics will be described and usage areas will be explained, developments and future roles will be mentioned and also expected achievements will be described. These achievements' relevance with learning outcomes of departments which are accredited by MUDEK will be defined.

---

[1] Araştırma Görevlisi, Anadolu Üniversitesi, ahmetmuratturk@anadolu.edu.tr
[2] Yardımcı Doçent Doktor, Anadolu Üniversitesi, abilge@anadolu.edu.tr

## Giriş

Teknolojideki hızlı değişim günlük hayatta da pek çok kimsenin farkında olduğu bir olgudur. Teknolojik gelişmeler, hemen her alanda, devletleri ve şirketleri etkilemekle kalmayıp kitleleri ve kültürleri de değiştirmektedir. Bilgiyi üreten, üretilen verileri işleyen ve elde edilen gelişmelerin uygulamaya geçmesini sağlayan kurumlar rekabette öne geçmektedir. Bilginin teknolojiye dönüşümüyle oluşan bilgi teknolojileri dünyadaki sınırları ortadan kaldırmakta ve gerek insanların gerekse işletmelerin küresel ölçekte bilgiye erişerek iletişim kurmasına ve rekabet etmesine olanak vermektedir. Değişim o kadar büyük bir hızla gerçekleşmektedir ki, kısa süreler içerisinde ileri teknoloji olarak görülen ürünler bile çağın gerisinde kalabilmektedir. 90'ların ikinci yarısından itibaren web sayfaları ulaşılabilir olmaya başlamışken, 2000'lerin başlarında bir kurum için web sayfasına sahip olmak itibar açısından önemliydi. Günümüzde ise işletmelerin kendilerine özgü bir mobil uygulaması bulunması bile oldukça sıradan hale gelmiştir. Değişim sadece erişim araçları ile sınırlı kalmamış, erişilen bilgiler de farklılaşmaya uğramıştır. İlk internet sitelerinde iletişim, yayıncıdan ziyaretçiye doğru tek yönlü iken, 2004 yılından itibaren kullanılmaya başlayan Web 2.0 kavramı (Kim & Shin, 2015; O'reilly, 2007; Pazzani & Billsus, 2007) ile toplumsal iletişim siteleri, bloglar, wiki sayfaları gibi kullanıcıların ortaklaşa ve paylaşarak oluşturduğu bilgi havuzları ortaya çıkmıştır. Bu süreçte, yayınlama süreçleri katılımlı hale dönüşmüştür. Arkasından geliştirilen sosyal paylaşım siteleri ise oldukça popüler hale gelmiştir. Bu süreçte, mümkün kıldıkları ortamla yazılımda yeni fırsatlar yaratan akıllı telefonlar da günlük hayatın ayrılmaz bir parçası olmuştur. Web 3.0 teknolojileri de (Chang vd., 2009) gelişmeye halen devam etmektedir. Güncel teknolojilerin bu hızlı değişimi bilgi teknolojileri eğitimlerini de hızla değişmeye zorlamaktadır.

Bütün bu adımların yanında pek çok insanın bilgisayarla olan etkileşimi yalnızca son kullanıcı düzeyindedir. Son kullanıcı, bilgi sistemlerinin arka planda hangi teknolojileri kullandığı, ne tür algoritmalar yürüttüğü ile ilgilenmez. Ancak, bir bilgi sistemi sadece başarım ile değerlendirilmez. 1990'lı yıllarda bilgisayar sistemlerindeki çalışmalar taşınabilirlik, doğruluk, çalışma süresi, bakım kolaylığı, işlevsellik, verimlilik gibi metrikleri iyileştirmeye yoğunlaşmışken, 2000'li yıllarla güvenlik, sürdürülebilirlik, ölçeklenebilirlik, kişisel gizlilik gibi konular önemli ölçütler haline gelmiştir. Görüldüğü gibi bir bilgisayar

120

sisteminde parametreler zamanla değişmekte, farklı kavramlar ön plana çıkmaktadır. Bütün bunların yanında 90'ların başında tartışılmaya başlanan kullanım kolaylığı da bir sistemin kullanıcılar tarafından kabul edilmesi için en önemli özelliklerden biridir (Davis, 1989). Bu ölçütlerin hepsini aynı anda iyileştirmek zor bir problemdir. Bir ölçütteki iyileştirme, diğer ölçütlerden taviz verilmesini gerektirebilir. Bu yüzden farklı problemlerin çözümlerinde optimizasyon sağlanarak sistem tasarımının yapılması gerekir. İhtiyaç duyulan bilgisayar sistemlerinin yapısı, tasarımı, geliştirilmesi ve sistemlerin kullanılması, gerekli yazılımların geliştirilmesi, donanım ve yazılım problemlerinin çözülmesinde bilgisayar mühendislerine ihtiyaç duyulmaktadır. Bilgisayar mühendisleri, hem bulunduğu zamanın kıstaslarına uygun sistemler tasarlamalı, hem de güncel teknolojileri kullanmalıdır. Yetkin bilgisayar mühendislerinin yetiştirilmesi, gelecek teknolojilerinde söz sahibi olmak için önem arz etmektedir. Bilgisayar mühendisliği programları da bu nitelikteki insan gücünü yetiştirmelidir.

Bu çalışmada, öncelikle bilgisayar mühendisliği eğitiminin bileşenleri incelenecek, daha sonra teknoloji eğilimlerine uygun ders içerikleri önerilecek, sonraki bölümde önerilen derslerin kabul görmüş öğrenme çıktıları ile ilişkileri belirtilecektir.

## Mesleki Eğitim

Bilgisayar Mühendisliği Eğitimini üç farklı alt alanda incelemek gerekmektedir. Bunlar mühendislik eğitimi, bilgisayar mühendisliği eğitimi ve bilgisayar bilimleri eğitimidir. Mühendislik eğitimi, alanı fark etmeksizin mühendis unvanı taşıyan kişilerden beklenen analitik düşünme, sentez yapabilme, sorunlara karşı en uygun çözümü bulabilme, araştırma yapabilme gibi yetileri kazanmasını sağlamak için verilen eğitimdir. Bu eğitim Matematik, Fizik, Kimya, Ayrık Matematik, İstatistik, Sayısal Analiz gibi temel bilim dersleriyle verildiği gibi, alan derslerinin kazanımları arasında da yer almaktadır. Bilgisayar mühendisliği ve bilgisayar bilimler farkı kavramlar olarak değerlendirilmelidir (Klump, 2013). Bilgisayar bilimleri verilerin ve komutların nasıl işlenmesi, saklanması ve diğer cihazlarla iletişim kurması hakkında yapılan çalışmalardır. Bilgisayar bilimleri algoritmaları ele alır. Kimi görüşler tarafından bir bilim dalı olup olmadığı tartışılmakta olsa da bilgisayar bilimlerinin bilgisayarla iç içe olmasının sebebi problemlerin çözümünde kullanılan kimi algoritmaların yüksek seviyeli hesaplama gerektirmesidir. Makine öğrenmesi, veri madenciliği, paralel programlama, hesaplama teorisi, kodlama teorisi, dağıtık hesaplama, evrimsel hesaplama, veri tabanı yönetim sistemleri bilgisayar bilimlerinin çeşitli alt dallarına örnek verilebilir. Bilgisayar mühendisliği ise mikroişlemcilerden akıllı telefonlara, dizüstü ve masaüstü bilgisayarlardan süper bilgisayarlara kadar çeşitli platformlar için tüm platformlarda bilgisayar bilimleri hesaplamalarının yapılması, platformların elektronik ve donanım olarak tasarımı ve optimizasyonu, verilerin elektronik bileşenlerin arasında paylaşılması, yazılımların geliştirilmesi, derlenmesi ve farklı platformları desteklemesi gibi konularla ilgilenir. Bilgisayar mühendisliğinin çalışma alanlarına örnek olarak sinyal analizi, yapay görü, doğal dil işleme, bilgisayar mimarisi, sistem programlama ve gerçek zamanlı hesaplama verilebilir. Bunların yanında bilgisayar mühendisliği eğitimi, bilgisayar ağlarını da kapsayarak ağ sistemleri, bulut hesaplama, ağ güvenliği, yönlendirme ve telekomünikasyon gibi konuları da kapsamaktadır.

Mühendislik eğitiminde kalite standartlarını denetleyen çeşitli bağımsız kuruluşlar vardır. Bu kuruluşlar, başvuruda bulunan bölümleri denetleyerek uygun gördükleri takdirde akredite etmektedir. Bu kuruluşlara örnek olarak Amerika Birleşik Devletleri merkezli faaliyet gösteren Accreditation Board for Engineering and Technology (ABET) ve

Türkiye'de faaliyet gösteren Mühendislik Eğitim Programları Değerlendirme ve Akreditasyon Derneği (MÜDEK) gösterilebilir. Akreditasyon kuruluşlarının değerlendirme raporlarında müfredatlar da incelenmektedir. Buradan hareketle güncel teknolojileri kapsayan bir müfredatın, akredite olma koşulu olarak değerlendirildiği söylenebilir.

## Teknoloji Eğilimleri

Bilgisayar bilimleri ve mühendisliğinin tarihsel gelişimi içerisinde ilgilendiği alanlar programlama dilleri, derleyiciler, işletim sistemleri, algoritmalar ve veri tabanları (Hopcroft, Soundarajan, & Wang, 2011) iken günümüzde internetin yaygınlaşması, üretilen veri miktarının artması, akıllı telefonların yaygınlaşması, sosyal ağların kullanımının artması gibi nedenler çözülmesi gereken yeni sorunları gündeme getirmektedir. Yüksek boyutlu verileri yönetme ve boyut indirgeme, büyük veri setlerini işleme, yapısal olmayan verilerden bilgi çıkarma ve sosyal ağ analizi günümüz koşullarında bilgisayar mühendislerinin öncelikli çalışma alanlarındandır.

### Büyük Veri

2012 yılı rakamlarına göre günlük 2.5 exabyte veri üretildiği, 2 zettabyte yayın yapıldığı düşünülmektedir ve 20 milyar web sayfasının yayında olduğu tahmin edilmektedir (Hopcroft, 2012; Hopcroft vd., 2011). Sosyal ağların popülerliği ve bu ağlardaki insanların etkileşimleri, akıllı sensörlerin ulaşılabilirliğinin artması ve maliyetinin düşmesi sonucu pek çok farklı amaç için kullanımının artarak fiziksel dünyada veri toplayıp depolamaları, GPS kullanan veya belirli bir erişim noktasına bağlanan konum duyarlı cihazlardan elde edilen verilerin saklanması, bu büyük verilerin elde edilmesini sağlar. Veriler metin, sensör bilgileri, ses, video ya da alışveriş fişleri gibi işlemsel bilgiler olabilir. Büyük veri sadece boyut olarak büyük değil, geleneksel yöntem ve araçlarla işlenemeyen veri anlamına gelmektedir (Zikopoulos & Eaton, 2011). Depolanmış verilerden çıkartılan bilgiler karar destek sistemlerine bilgi sağlanması ile hızlı şekilde sonuç üretilerek harekete geçilmesini sağlayacak platformlar bütününe büyük veri sistemleri denilmektedir. Ham veriler işlenerek anlamlı hale getirilir ve kişiye özel kampanyalar üretilip daha fazla müşteriye ulaşma sağlanır, bir malın en uygun fiyatlandırması sağlanır, konum verileriyle hangi saatlerde nerelerde yoğunluk olduğunu tespit edilip yeni mağaza açma ya da marketler için tedarik sağlanır. Finans sektöründe riski en aza indirecek tahminler yapılması sağlanır. Buna benzer pek çok farklı alanda büyük verinin kullanım alanını örneklendirmek mümkündür. Yapılan bir çalışma dünyadaki verilerin %80'inden fazlasının korunmadığını ve elde edilen verilerin sadece %0,5'inin analiz edildiğini göstermektedir (Gantz & Reinsel, 2012). Hadoop, JAQL, HBASE, NOSQL kavramları büyük verinin ilgi alanındaki teknolojilerdir (Loganathan, Sinha, Muthuramakrishnan, & Natarajan, 2014). Hadoop pek çok teknoloji devi tarafından kullanılan ve geliştirilen bir teknoloji olmasının yanı sıra açık kaynak kodlu olduğu için, müfredata eklenmesinde bir engel yoktur. Yetiştirilecek mühendislerin gerek güncel bilgi teknolojilerine uyum sağlaması, gerekse büyük veri alanının sağladığı iş imkânları nedeniyle müfredata eklenmesi gerekmektedir. Columbia Üniversitesi gibi bazı üniversiteler lisansüstü seviyesinde dersler vermekte, pek çok eğitim kurumu da sertifika programları ile büyük veri eğitimi vermektedir.

### Nesnelerin İnterneti ve Giyilebilir Teknolojiler

Nesnelerin interneti günlük hayatta kullanımda olan nesnelerin internet ağına bağlanarak veri gönderip alması kabiliyeti olarak tanımlanmaktadır (Weber & Weber, 2010).

122

Nesne kavramı oldukça geniş bir anlama sahip olduğu için bu alan pek çok farklı donanımları kapsamaktadır. Akıllı Şehir Uygulamaları, Akıllı Çevre Uygulamaları, Akıllı Ev Uygulamaları, Akıllı Hayvancılık ve Tarım Uygulamaları, Perakende Tedarik Uygulamaları, e-Sağlık Uygulamaları, Endüstriyel Kontrol, Lojistik, Akıllı Ölçmeler gibi (Atzori, Iera, & Morabito, 2010; 50 Sensor Applications for a Smarter World) alanlarda nesnelerin internetinin kullanıldığı düşünülürse, konunun kapsamının ne kadar geniş olduğu anlaşılabilir. Nesnelerin interneti konusu aynı zamanda başka bir popüler konu olan bulut bilişimle de yakından ilişkilidir (Mitton, Papavassiliou, Puliafito, & Trivedi, 2012). Nesnelerin interneti kavramı ilk olarak 1999 yılındaki yapılan bir sunumda kullanılmış ve son yıllarda internet taşınabilirliğinin ve ulaşılabilirliğinin gelişmesiyle popüler hale gelmiş bir konudur (Gubbi, Buyya, Marusic, & Palaniswami, 2013). Nesnelerin internetinin pazara uyumu için 10 yıl kadar bir zaman öngörülmüştür. Arabalar, elektrikli aletler, yiyecekler, buzdolapları, su ısıtıcıları, akıllı yapılar (bina, sera, besihane vb.) farklı cihazlara bağlanarak makine-makine bağlantıyı oluştururlar. Farklı alanlar için sistemler geliştirilirken kullanıcı ve sistem ara yüzü tasarımı, veri tabanı tasarımı, gömülü sistem tasarımı ve yazılım tasarımına ihtiyaç duyulmaktadır. Nesnelerin internetini konu alan bir derste, sistem tasarımının nasıl yapılması gerektiği üzerinde durulmalıdır. Tasarımlar uygulamaya özgü ve ihtiyaca yönelik oldukları için, bilgisayar mühendislerinin senaryoya göre en uygun çözümü bulmalarına yönelik çalışmalar yapılmalıdır. Örneğin internete erişim hızının düşük olduğu ve servis sağlayıcıların koydukları kullanım kotasının olduğu bir sistem için tasarlanacak veri paketlerinin boyutu mümkün olduğunca küçük olmalı, gönderilen veriler veri tabanı tasarımıyla ilişkilendirip daha kapsamlı verileri elde etmeye yönelik olmalıdır. Bu amaca uygun olarak Şekil 1'de görülen elektronik kartla yapılan bir çalışmada internet bağlantısı ve elektrik şebekesine erişimi kısıtlı bir alan için mobil bağlantı kullanan akıllı sera otomasyonu önerilmiştir (Turk, Sora Gunal, & Gurel, 2015). Böyle bir tasarımı gerçekleştirmek için bir bilgisayar mühendisliği öğrencisinin veri tabanı, mikroişlemciler, bilgisayar ağları, mobil uygulama geliştirme ve web sayfası geliştirme konularında bilgi sahibi olması gerekmektedir. Bu sebeple bu içerikteki bir ders, son sınıf öğrencilerine hitap edecektir.

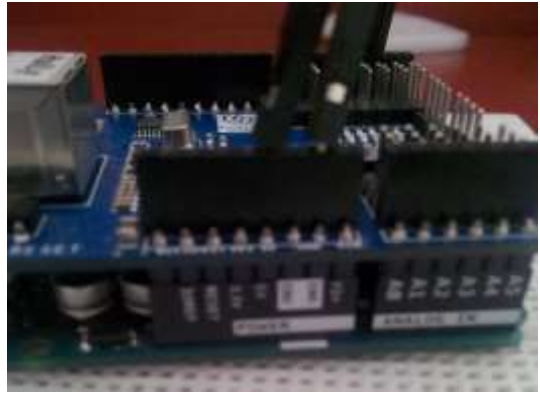

Şekil 1. İnternet kontrollü gömülü sistem kartı

Giyilebilir Teknolojiler, nesnelerin internetinin bir alt dalı olabileceği gibi, başlı başına farklı bir çalışma alanı olarak da incelenebilir. Akıllı saatler, interaktif ve akıllı gözlükler, akıllı bileklikler bu alanda var olan mevcut başlıca ürün kategorileridir. Her bir giyilebilir cihaz için tasarım problemleri tartışılmalı ve kabul edilebilir bir model aranmalıdır. Sağlık alanında da giyilebilir teknolojiler bir süredir kullanılmaktadır (Starner, Auxier, Ashbrook, & Gandy, 2000). Kullanım amacına uygun protokolleri kullanan giyilebilir cihazların tasarımı

123

ve kullanımı hakkında bilgisayar mühendisliği öğrencilerine temel bilgiler aktarılmalı, öngörü kazandırılmalıdır.

**Öneri Sistemleri**

Öneri sistemleri yeni bir kavram olmasa da, bilginin aşırı miktarda artmasına karşılık getirdiği filtreleme çözümü ile e-ticaretin, çevrimiçi izleme-okuma sistemlerinin yaygınlaşmasıyla popüler bir çalışma alanı olmuştur. Film, kitap, müzik, otel, lokanta ve sosyal hizmet başta olmak üzere çeşitli alanlarda veri filtreleme amaçlı kullanılan öneri sistemlerine, Facebook'ta "Tanıyor Olabileceğiniz Kişiler", Twitter'da "Kimi Takip Etmeli" önerileri örnek verilebilir. Öneriler temel olarak iki farklı strateji ile yapılır. Bunlardan ilki olan içerik tabanlı öneri sistemleri bir ürüne benzeyen diğer ürünleri çeşitli metotlarla keşfeder. Bir kullanıcının ilgilendiği ürünün benzerine de yine kullanıcı tarafından ilgi duyulacağı varsayılır. E-ticaret sitelerinde Şekil 2'deki gibi, bir ürünle ilgilenirken sayfada çıkan "Bu Ürüne Bakanlar, Şu Ürünlere de Baktılar" şeklindeki öneriler, içerik tabanlı öneri sistemlerine bir örnektir. Market sepetlerindeki birliktelik analizine dayanan "Bu Ürünü Alanlar, Bununla Birlikte Şu Ürünü de Aldılar" önerileri de içerik tabanlı öneri sistemlerine örnek verilebilir.

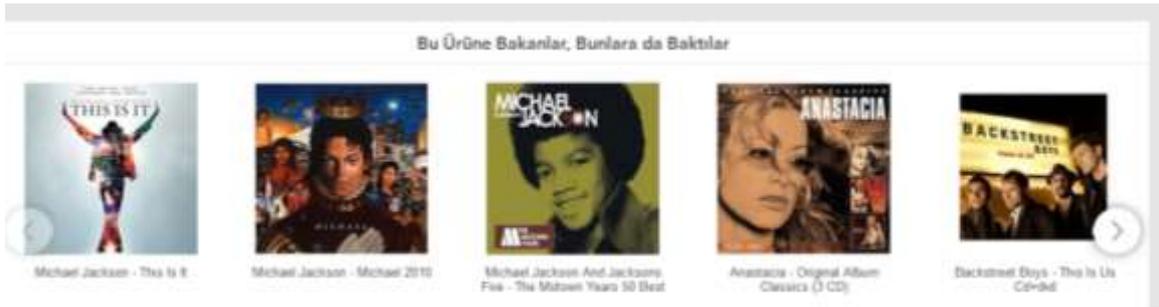

Şekil 2. İçerik tabanlı öneri sistemi (www.hepsiburada.com sitesinden alınmıştır.)

Şekil 3'de bir kesiti görünen ve internet devi Amazon'a ait olan, film veri tabanı oluşturan, dünyadaki tüm filmler hakkında bilgiler saklayan imdb.com, kullanıcılarından her film için 1-10 tercih aralığında bir puanlama yapmasını istemektedir. Bu puanları kullanarak, bir filmin tanıtım sayfasında "Bu Filmi Seven Kişiler, Aynı Zamanda Şu Filmleri de Sevdiler" şeklinde bir öneri sunmaktadır.

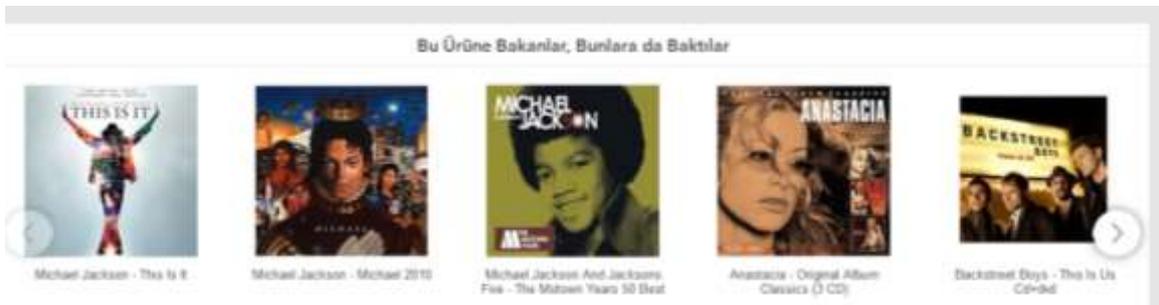

Şekil 3. Tercih puanına dayalı içerik tabanlı öneri sistemi

Diğer bir öneri stratejisi ise ortak filtrelemedir. Ortak filtreme sistemleri imdb.com'a benzer şekilde kullanıcıdan bir ürün hakkında puanlama yapmasını beklemekte, her bir ürüne

124

verilen skorları kullanarak, puanlamayı yapan kullanıcıya çıktı olarak en seveceği N tane ürünü önerir veya herhangi bir ürünü ne derecede seveceğini tahmin eder. Ortak filtrelemenin temel varsayımı geçmişte benzer davranışları sergileyen kimseler, gelecekte de benzer davranışları sergilemeye eğilimlidir varsayımına dayanır (Herlocker, Konstan, Borchers, & Riedl, 1999). Böylelikle bir kullanıcının, hiç izlemediği bir filmi izlemiş olsa hangi derecede beğeneceği tahmin edilir. Ortak filtreleme algoritmalarının doğruluğunu arttırma ve ölçeklenebilirliğini sağlama (Bell & Koren, 2007), gizliliği koruyarak öneri üretme (Polat & Du, 2005) gibi çeşitli araştırma alanları gelişimini devam ettirmektedir. Bunun yanında öneri sistemleri manipüle edilmeye açıktırlar (Gunes, Kaleli, Bilge, & Polat, 2014).

Bir bilgisayar mühendisliği öğrencisinin mevcut veri yapısı ve kalitesiyle fazla miktardaki veri içerisinden kendisinden tavsiye isteyen kullanıcı için yararlı bilgileri sunabilecek uygun bir öneri sistemi tasarlaması için müfredata bu içeriklerin eklenmesi gerekmektedir. Çevrimiçi açık kurs sistemi Coursera'da (www.coursera.org) "Öneri Sistemlerine Giriş" kursu vardır. Kullanıcıların mobil telefonlardan beklentileri arttıkça, sistemden konum bilgisini kullanarak önerilerde bulunması beklenmektedir.

**Bulut Teknolojileri**

Yapılandırılabilir paylaşılmış bilişim kaynakları havuzlarına (sunucular, depolama alanları, ağlar, uygulamalar ve servisler gibi) talep olduğu anda, her yerden elverişli şekilde ağ erişimi sağlayan, kaynakların ihtiyaç anında hızlı bir şekilde eklenip, kullanılmayacağı zaman kolayca çıkartılmasına olanak verecek şekilde yönetilebilen modele bulut bilişim adı verilmektedir (Mell & Grance, 2009). Sunucu havuzu paylaşımına örnek olarak şu durum verilebilir: bilimsel bir deney yapmak isteyen akademisyen yüksek hesaplama hızına ihtiyaç duyar, bu amaçla satın alınacak bir süper bilgisayar yüksek yatırım maliyeti gerektirir. Bunun yanında deney sonunda süper bilgisayara ihtiyaç duyulmayacağı için kaynaklar verimli şekilde kullanılmamış olur. Benzer şekilde bir e-ticaret sitesi, yılın belirli dönemlerinde (bayramlar, yılbaşı, sevgililer günü vb.) daha fazla ziyaretçiye hizmet verebilmek için altyapısını güçlendirmelidir. Ancak bu dönemler dışında, e-ticaret şirketinin kullanmadığı kaynakları ek maliyet getirmektedir. Bulut altyapısı kullanan bir e-ticaret şirketi, ihtiyaç anında kaynak ekleyerek ve yoğun dönem bittikten sonra ek kaynakları çıkartarak en uygun maliyetle yayın yapabilir duruma gelebilir. Bunun yanında sağladığı dosya ve veri paylaşım havuzuyla da bir dosyayı özel, halka açık ya da belirli bir grup için erişilebilir olarak sunabilir. Örneğin şirket içi bir projede, sadece projeye dâhil olan çalışanların ilgili dosyalara erişip değiştirme yetkisi vererek proje yönetimi kolaylaştırılabilir. Aynı zamanda verilere yetkili kişilerin şirket içinden, şirket dışından ve mobil olarak ulaşması sağlanabilir. Benzer bir örnek olarak hastaların tıbbi geçmişlerinin bulut ortamında saklanarak, hasta muayene olacağı anda doktorun kayıtlara ulaşarak teşhis koymasını kolaylaştırıcı bilgilere kolayca ulaşması sağlanabilir. Sağladığı bu avantajın yanında gizlilik içeren bu verilerin buluttaki güvenliği sorgulanmaktadır (Zissis & Lekkas, 2012). Pek çok kullanım alanı içeren bulut teknolojilerinin kullanım alanı giderek yaygınlaştığı için bu teknolojilerin altyapıları, mimarileri, programlanmaları hakkında bilgi sahibi olan bilgisayar mühendisi yetiştirmek çağın gereksinimlerinden biridir.

**Kimlik Yönetimi ve Siber Güvenlik**

Kimlik yönetimi, sistemdeki kullanıcıların hangi sistemlere ve hangi verilere ulaşabileceğini denetleyen mekanizmadır. Gerek büyük ölçekli kurumlarda gerekse çapraz kimlik doğrulaması gerektiren sistemlerde kimlik yönetim protokolleri kullanılmalıdır. Büyük

125

ölçekli yazılımlar, kurumsal şirketler için kimlik yönetimi çözümleri sunulmaktadır, ancak sistemler ne kadar güvenilir olursa olsun, sistemlerin en zayıf noktası sistemi kullanan insanlar olarak değerlendirilmektedir (Gollmann, Herley, Koenig, Pieters, & Sasse, 2015). İstatistiklere göre başarılı saldırıların büyük bir kısmı kullanıcı kimliği çalınarak gerçekleşmektedir (Cyber crime: Attacks experienced by U.S. companies 2015). Güvenilirliği yüksek platformlarda bile sosyal mühendislik metotları kullanan saldırgan, çeşitli senaryolarla kullanıcılara istediğini yaptırmakta ve sistemlere zarar vermektedir. Günümüzde savaşların teknoloji ile yapıldığı ve çok önemli bilgilerin bile dijital ortamda saklanıldığı düşünüldüğünde siber güvenlik dersi bilgisayar mühendisliği programları müfredatına dâhil edilmelidir.

## Öğrenme Çıktıları

MÜDEK tarafından akredite edilen bilgisayar mühendisliği bölümleri web sayfaları incelenmiştir. Anadolu Üniversitesi, Ankara Üniversitesi, İzmir Ekonomi Üniversitesi, Osmangazi Üniversitesi, Yeditepe Üniversitesi gibi bazı bölümlerin MÜDEK tarafından önerilen öğrenim çıktılarını disipline özgü çıktı olarak şu şekilde önerildiği görülmüştür.

(01) Matematik, fen ve Bilgisayar mühendisliği konularında yeterli bilgiye sahip olma ve bu bilgileri, alanındaki problemlerin çözümü için kullanabilme becerisi

(02) Bilgisayar mühendisliği problemlerini uygun analiz ve modelleme yöntemleri kullanarak saptama, tanımlama, formüle etme ve çözme becerisi

(03) Bir sistemi, bileşeni veya süreci, gereksinimleri karşılamak üzere gerçekçi kısıtlar altında, modern yöntemler kullanarak tasarlama becerisi

(04) Bilgisayar mühendisliği uygulamaları için gerekli olan modern teknik ve araçları geliştirme, seçme, kullanma ve bilişim teknolojilerinden etkin bir şekilde faydalanma becerisi

(05) Deney tasarlama, deney yapma, veri toplama, sonuçları analiz etme ve yorumlama becerisi

(06) Disiplin içi takımlarda, çok disiplinli takımlarda ve bireysel çalışabilme becerisi

(07) Türkçe ve İngilizce olarak sözlü ve yazılı etkin iletişim kurma becerisi

(08) Yaşam boyu öğrenme bilinci, bilgiye erişebilme, bilim ve teknolojideki gelişmeleri izleme ve kendini sürekli yenileme becerisi

(09) Mesleki ve etik sorumluluk bilinci

(10) Proje yönetimi, risk yönetimi ve değişiklik yönetimi gibi iş hayatı uygulamaları hakkında bilgi; girişimcilik, yenilikçilik ve sürdürülebilir kalkınma hakkında farkındalık

(11) Bilgisayar mühendisliği çözümlerinin evrensel ve toplumsal boyutlarda sağlık, çevre ve güvenlik üzerindeki etkileri, hukuksal sonuçları ve çağın sorunları hakkında bilgi

Müfredata eklenilmesi önerilen derslerin öğrenim çıktıları ile ilişkisini ifade etmede kullanılan notasyon ise şu şekildedir.

0: Katkısı Yok

1: Giriş
2: Pekiştirme
3: Geliştirme (Vurgulama)

|  | PÇ-1 | PÇ-2 | PÇ-3 | PÇ-4 | PÇ-5 | PÇ-6 | PÇ-8 | PÇ-9 | PÇ-10 | PÇ-11 |
|---|---|---|---|---|---|---|---|---|---|---|
| Büyük Veri | 3 | 2 | 1 | 3 | 2 | 2 | 3 | 2 | 1 | 0 |
| Nesnelerin İnterneti | 3 | 3 | 3 | 3 | 2 | 2 | 3 | 2 | 3 | 2 |
| Öneri Sistemleri | 3 | 3 | 2 | 2 | 2 | 2 | 3 | 2 | 2 | 2 |
| Bulut Teknolojileri | 3 | 2 | 1 | 3 | 1 | 2 | 3 | 2 | 1 | 1 |
| Kimlik Yönetimi ve Siber Güvenlik | 3 | 3 | 3 | 1 | 3 | 2 | 3 | 2 | 2 | 3 |

Tablo 1. Programların Öğrenme Çıktıları ile Önerilen Derslerin İlişkileri

## Sonuç

MÜDEK tarafından akredite edilen bilgisayar mühendisliği bölümlerin ders programları incelendiğinde bölümlerin mobil teknolojiler, paralel programlama, uzman sistemler gibi günümüzde ihtiyaç duyulan bilişim sistemlerinin temellerini oluşturan konulara müfredatlarında yer verdikleri görülmüştür. Her bölümün ağırlık verdikleri programlama dillerinin de farklı olduğu görülmüştür. Birden fazla programlama dili bilmek bir mühendis için olmazsa olmazıdır ancak eğitim hayatında ağırlık verilen dilin öğrenci tarafından daha fazla tercih edileceği de bir gerçektir. Bundan dolayı müfredat oluştururken dünya çapında kabul görmüş ve eğiliminin fazla olduğu dillere ağırlık verilmelidir. Sırasıyla Java, C, C++, Phyton ve C# dilleri dünyada en çok kullanılan dillerdir (Cass, S). Üniversitelerde eğitime Java diliyle başlandığı gözlenmiştir, bu karar piyasa eğilimlerini takip etmek açısından olumlu bir gelişme olarak değerlendirilebilir. Yine de bazı üniversitelerde ders içerikleri sınıf nesne kavramları ile başlamaktadır. Mesleki eğitime yeni başlayan öğrenciler için anlaşılması zor konular verilmemelidir. Giriş aşamasında nesne yönelimli dil kavramları yerine, yapısal dil kavramlarından bahsedilmelidir. Java programlama dili web, masaüstü ve mobil platformlarda kullanılabildiği için, Java kullanan yazılımcılara büyük avantajlar sağlamaktadır. Ancak diğer diller de ihmal edilmemeli, araştırma grupları, bilgisayar kulüpleri ve geliştirici gruplarına ders dışı aktivitelere bu konularla ilgili seminer, hackathon, yarışma gibi aktivitelere teşvik edilmelidir. Son zamanlarda çeşitli şirketlerin ve geliştirici grupların etkinlik düzenledikleri gözlenmektedir. Bu etkinliklerde güncel teknolojilerin, özellikle mobil yazılım geliştirmeye yönelik oturumlarının olduğu gözlenebilir. Müfredatta yer alamayacak kadar detaylı ve hakkında bilgi bulmanın zor olduğu özel oturumlarla öğrencilerin kişisel gelişimi sağlanmalıdır. Ayrıca müfredatın sektör ile ilişkisini arttırmak için mesleki derslerde konuyla ilgili özel sektörden destek alınmalı, çeşitli şirketlerden davetli konuşmacılarla dersler desteklenmelidir.

Bulut teknolojiler, giyilebilir cihazların üretimi ve nesnelerin interneti konuları büyük ölçekli şirketlerin radarına giren çalışma alanlarıdır. Öğrencilerin bu konularda, ders

127

materyallerinin yanına konu ile ilgili uzaktan eğitim materyalleri de kullanılarak ders dışı aktiviteler ile gelişimleri desteklenmeli ufuklarının açılması sağlanmalıdır. Önerilen teknoloji eğilimlerinin olduğu dersler proje tabanlı öğrenme (Dym vd. 2005) ile müfredata eklenmelidir.

**Kaynakça**